\documentclass[aps,prd,showpacs,eqsecnum]{revtex4}
\usepackage{amsmath,amssymb}
\usepackage{graphicx}
\newcommand{\vect}[1]{\!\!\!\mbox{ \boldmath $#1$}}

\begin{document}

\title {Gravitational wave signals from a chaotic system:
A point mass with disk}

\author{Kenta Kiuchi$^{1}$
\footnote{\affiliation\ kiuchi@gravity.phys.waseda.ac.jp}}
\author{Hiroko Koyama$^{1}$
\footnote{\affiliation\ hiroko.koyama@gravity.phys.waseda.ac.jp}}
\author{Kei-ichi Maeda$^{1,2,3}$
\footnote{\affiliation\ maeda@waseda.jp}}
\affiliation{$^{1}$Department of Physics, Waseda University, 3-4-1 Okubo,
 Shinjuku-ku, Tokyo 169-8555, Japan~}
\affiliation{$^{2}$Advanced Research Institute for Science and Engineering,
 Waseda University, Shinjuku, Tokyo 169-8555, Japan}
\affiliation{$^{3}$Waseda Institute for Astrophysics,
 Waseda University, Shinjuku, Tokyo 169-8555, Japan}

\date{\today}

\begin{abstract}
We study gravitational waves from a particle moving around a system of a 
point mass with  a 
disk in Newtonian gravitational theory. A particle motion 
in this system can be chaotic when the gravitational contribution 
from a surface density of a disk is comparable with that 
from a point mass.
In such an orbit, we sometimes find that there
appears a phase in which 
particle motion becomes nearly regular 
(so-called ``stagnant motion'' or ``stickiness")
for a finite time interval between more strongly 
chaotic phases. 
To study
how these different chaotic behaviours affect 
observation of gravitational waves,
 we investigate a correlation of the 
particle motion and the waves. 
We find that 
such a difference in chaotic motions
 reflects on the wave forms and energy spectra.
The character of the waves in the stagnant motion
is quite different from that either in a regular motion
or in a more strongly chaotic motion.
This suggests that we may make a distinction 
between different chaotic behaviours of the orbit
via the gravitational waves.
\end{abstract}


\maketitle

\baselineskip 18pt
\section{Introduction}
Chaos appears universally in nature and it is expected to be a fundamental tool
to understand various nonlinear phenomena. 
Historically, research on chaos started 
from the famous three-body problem by Poincar\'{e},
 and so far many attempts to reveal the character of 
chaos have been done in dynamical systems. Following 
such works, considerable 
research in Newtonian gravity and general relativity have been 
done~\cite{Hobill,Barrow,Conto,Karas,Varvoglis,Bombelli,Moeckel,Dettmann,Yust,
Sota,Suzuki,letelier_viera,podolsky_vesely,Levin,Schnittman,Cornish,
Kiuchi:2004bv,Koyama:2007cj,Karney:1995,Contopoulos:2000,Contopoulos:2002}.
But most of it, especially work on relativistic systems, 
has discussed only whether  or not
chaos occurs mainly by using 
 the Poincar\'{e} map and the Lyapunov exponent. 
However, we know there appear various types of chaotic behaviours
depending on the strength of chaos and its analysis will
play a very important role to understand the essence of nonlinear dynamics
~\cite{Karney:1995,Contopoulos:2000,Contopoulos:2002}. 
A coquet view of how one can extract and use information from a chaotic system 
may be also missing. So one can address
 two new important issues in the research of chaos
 in Newtonian gravity  and general relativity.
One is, of course, to make clear the character of chaos for each system,
and the other is to find some methods 
to extract useful information from chaotic systems. As for the first point, 
we have recently shown 
the possibility to classify the character of
chaos in a system of a
 spinning particle moving around a Schwarzschild black hole
\cite{Koyama:2007cj}.
The method used in~\cite{Koyama:2007cj}
is a power spectrum analysis of 
the particle orbit. The spectra are mainly classified into 
a power-law type and a white-noise type. As a result, we find that 
there is a close
relation between the so-called  ``stagnant motion''
(or 
``stickiness"\cite{Karney:1995,Contopoulos:2000,Contopoulos:2002}) 
and a ``power-law'' spectrum.

As for the second point, an indirect method to extract
 information from chaotic systems 
is required for the following reason: In chaotic systems in astrophysics, 
it is sometimes too difficult to observe chaotic motions directly. 
Because these systems are often far from the earth and 
the ambient surroundings of  
these systems may not be clean. Therefore, in this paper
 we propose the use of gravitational 
waves as a new method to analyze chaos. 
The reason we choose gravitational waves
is as follows: 
Detection of gravitational waves is one of the greatest challenges 
in experimental and theoretical physics in this decade. Several 
kilometer-size laser interferometers, such as TAMA~\cite{Tsubono:1994sg}, 
LIGO~\cite{Abramovici:1992ah}, and GEO~\cite{Hough:1996nx} are 
 now in operation. In addition to these ground-based detectors, the 
Laser Interferometer Space Antenna (LISA) with an arm length of 
$5 \times 10^6$km has been proposed and is planned to start 
observation in the near future~\cite{Thorne:1995xs}. 
Gravitational waves will 
bring us various new information about
relativistic astrophysical objects. If we detect gravitational waves
and compare them with theoretical templates, we may be able to determine a
variety of astrophysical parameters of the sources such as their
direction, distance, masses, spins, and so on. 
The direct observation of
gravitational waves could resolve strong-gravitational phenomena such as
a black hole formation.
Furthermore, we may be able not only to verify the theory of
gravity but also to find new information  at 
high density or to recover new physics  in a high energy region.

In \cite{Kiuchi:2004bv}, we analyzed the gravitational waves
from a spinning test particle in a Kerr black hole.
We find that there is a difference between the spectra of
the gravitational waves from a chaotic orbit
and from a regular one.  There appear many small spikes in the spectrum
of the chaotic orbit.
However, as we mentioned, there are various types of chaotic motions,
and it is important in the analysis of such a
dynamical system to know which type of chaos appears
as well as to show the difference between a regular motion and a chaotic one.
Hence,
in order to study whether one can make a distinction between various types of
chaos by use of gravitational waves,
 we should reanalyze them in a chaotic system.

As a concrete model of a chaotic system, here we consider 
a point mass with a thick disk in Newtonian
 gravity~\cite{Saa:1999je,Saa:2000ec}.
 This model mimics a system of a black hole with a massive accretion
 disk\cite{footnote1}.
Saa analyzed this system and showed that
a particle motion is chaotic~\cite{Saa:2000ec}.
This model can describe almost regular to
highly chaotic motion by changing the ratio of a disk mass to a black hole 
mass.
In~\cite{Saa:2000ec}, however, 
only the Poincar\'{e} map has been analyzed to judge whether 
chaos occurs or not, 
and the characteristics of chaos have not been studied much.

So our strategy in this paper is the following: 
First, we analyze the particle motion and make clear the characteristics of 
chaos appearing in this system. Secondly, we evaluate 
the gravitational waves from such a system by use of  
the quadrupole formula. Finally, to study some observational feature of 
chaos appearing in the gravitational waves, we investigate correlation 
between types of chaotic  motions and gravitational waves, and
 then point out a possibility to extract information about this 
chaotic system from the gravitational waves.

This paper is organized as follows. In Sec.~\ref{basic}, 
we shall briefly summarize the basic equations. Numerical analysis 
results will be presented in Sec.~\ref{numerical}. The summary and
discussion follow in Sec.~\ref{summary}. Throughout this paper,
we use the geometrical units of $c=G=1$.

\section{Basic equations}\label{basic}

We start by considering the Newtonian limit of a black hole disk system~
\cite{Saa:1999je}.
The equations of motion for a test particle 
in this background are very simple.
We use the cylindrical coordinates $(\varpi,\varphi,z)$.

A point mass  with a mass 
$M$ is located at the origin, while a disk 
exists on the equatorial plane ($z=0$).
A smooth distribution of disk matter is assumed. 
If the radial gradient of the density 
is much smaller than vertical one,
we can approximate the density as $\rho=\rho(z)$. 
A minimal but realistic model 
for a rotating thick disk may be described by
 Emden's equation~\cite{Saslaw:1985}.
 As in~\cite{Saa:2000ec}, 
ignoring the radial gradient, we find that Emden's equation for 
disk matter density $\rho(z)$ is given by
\begin{eqnarray}
\kappa \gamma \rho^{\gamma-2}\frac{d^2 \rho}{dz^2} + 
\kappa \gamma ( \gamma - 2 )\rho^{\gamma-3}\left(
\frac{d \rho}{dz}\right)^2 
= - 4 \pi \rho\,, \label{eq2-3}
\end{eqnarray}
where $\kappa$ and 
$\gamma = 1 + 1/n $ are the polytropic constant and 
 the polytropic index, respectively. 
The matter density $\rho$ should obey 
the Poisson equation $\nabla^2 V_D = 4 \pi \rho$,
where $V_D$ is the potential of the disk.
 For the isothermal case $(\gamma = 1 )$, 
Eq.~(\ref{eq2-3}) has the analytic solution,
\begin{eqnarray}
\rho(z) = \frac{\alpha}{4\pi z_0}\text{sech}^2\left(\frac{z}{z_0}\right), 
\label{eq2-4}
\end{eqnarray}
which corresponds to the disk potential
\begin{eqnarray}
V_D(z) = \alpha z_0 \ln\cosh\left(\frac{z}{z_0}\right),\label{eq2-6}
\end{eqnarray}
where $z_0$ and $\alpha$ describe the ``thickness'' of a disk 
 and the surface mass density, respectively. 
These two parameters determine the polytropic constant by 
 the relation $2\kappa = \alpha z_0$. In the limit of 
$z_0 \to 0$, we recover the potential of an infinitesimally thin disk 
$(V_D\sim \alpha|z|)$. The corresponding matter distribution is
given by  the $\delta$ function from Eq.~(\ref{eq2-4}).

Thus, the dynamics of a test particle with a  mass $\mu$
moving around 
a system of a point mass with a
smooth thick isothermal disk will be governed by the following  
(effective) Hamiltonian
\begin{eqnarray}
H = \mu\left[
\frac{\dot{\varpi}^2}{2} + \frac{\dot{z}^2}{2} + \frac{L^2}{2\mu^2\varpi^2} 
- \frac{M}{\sqrt{\varpi^2+z^2}} 
+ \alpha z_0 \ln\cosh \left(\frac{z}{z_0}\right)
\right]
\,,
\label{eq2-5}
\end{eqnarray}
where $L$ is the angular momentum of a particle and the dot denotes
the time derivatives.
\section{Numerical Analysis}\label{numerical}
\subsection{Two phases of chaos in particle motion}
At first, we analyze particle motion.
We numerically integrate the equations of motion of a test particle. 
The symplectic scheme is used because we have the 
analytic form of the Hamiltonian in this system. 
The integrated time is enough long such that
a particle moves thousands times 
around the central mass.
The numerical accuracy is 
monitored by the conservation of the Hamiltonian, 
which is typically 
$10^{-8} \sim 10^{-9}$.
It guarantees that our numerical calculation is reliable.
\begin{figure}
\includegraphics[width=7.5cm]{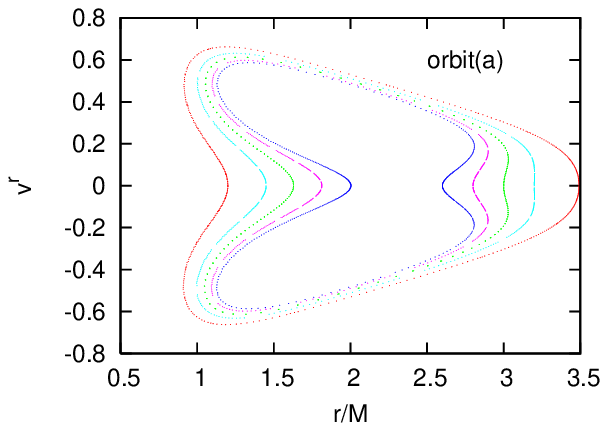}
\includegraphics[width=7.5cm]{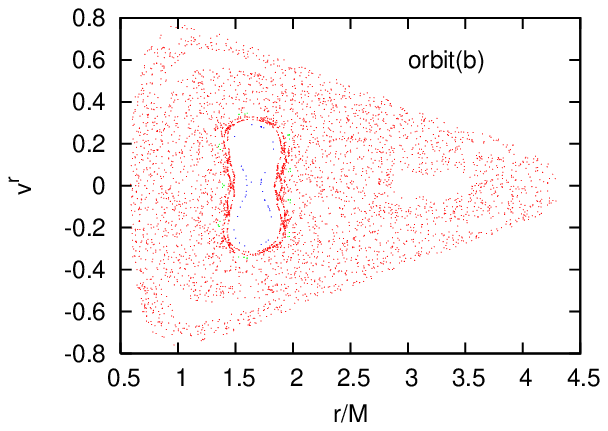}
\includegraphics[width=7.5cm]{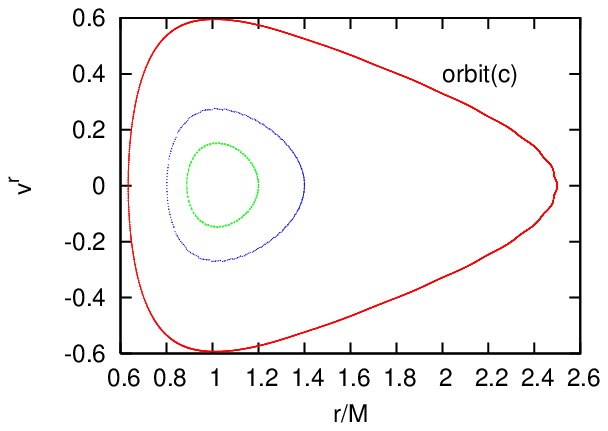}
\caption{\label{pmap-Newton} Poincar\'{e} maps 
of orbits of a particle with $ H = - 0.2$ and $ L = 1 $
across the plane 
$z=0$ in  a system of a point mass with
 a  disk. 
We set
the thickness the of the disk $z_0=0.5$, 
and its surface density 
(a) $\alpha =0.01$, (b) $\alpha = 0.1$, or (c) $\alpha = 10.0$. 
All figures are superpositions of trajectories starting from
 different initial conditions. In Figs.~(a) and (c), 
all trajectories form regular tori.  In Fig.~(b), 
 some trajectories from certain initial 
conditions  still seem to form tori, but others do not.
In fact, one initial condition, 
$(\varpi,v^\varpi,z,v^z)=(1.2,0,0,0.76)$ gives an 
almost two dimensional map on which the orbital points are widely
scattered, which means that
the particle motion is chaotic (We call it Orbit (b)). 
The outermost trajectories in Fig.~(a) 
and (c) are called Orbit (a) and (c), whose initial conditions are 
$(\varpi,v^\varpi,z,v^z)=(1.2,0,0,0.76)$ and $(2.5,0,0,0.49)$, respectively. 
}
\end{figure} 
\begin{figure}
\includegraphics[width=7.5cm]{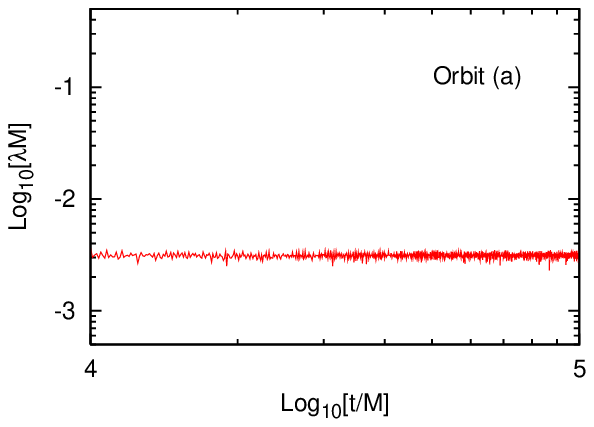}
\includegraphics[width=7.5cm]{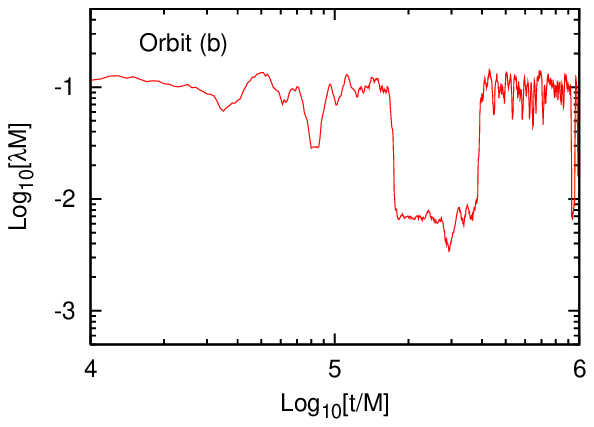}
\includegraphics[width=7.5cm]{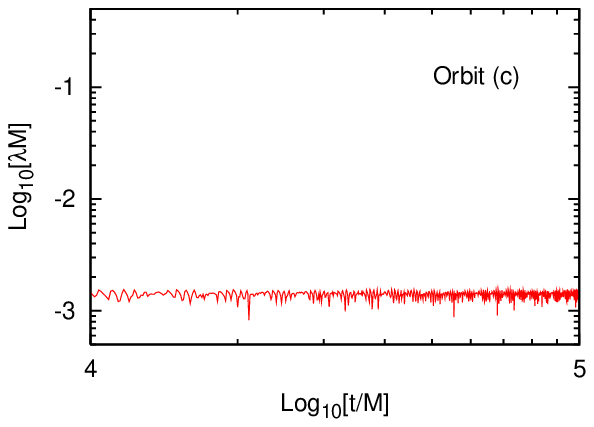}
\caption{\label{lyapu-Newton} The time evolutions of 
``local" 
Lyapunov exponents for 
Orbits (a), (b), and (c) in Fig.~\ref{pmap-Newton}. The Lyapunov 
exponents for Orbits (a) and (c) settle down to very small values, 
but that for Orbit  (b) is  large and changing in time.
It decreases to a very small value 
in the time interval of $t/M = (1.6  \sim 3.8)\times 10^5$.}
\end{figure}
We set $M=1$ to fix our units.
There are two parameters of a disk which we can change, i.e., 
the surface density $\alpha$ and the width $z_0$.
Which parameter dependence we should analyze ? 
When we change $\alpha$, there are two extreme limits, i.e., 
$\alpha \to 0$ and $\alpha \to \infty$, in which
 the system becomes integrable (see Eq.~(\ref{eq2-5})). 
The gravity by the central mass becomes
dominant when $\alpha \to 0$, while the force driven by the 
disk is dominant as $\alpha \to \infty$.
On the other hand, if we consider two extreme limits of $z_0$, i.e., 
the limits of $z_0 \to 0 $ and of $z_0 \to \infty$,
we find that  the system is still nonintegrable even in such limits~
\cite{Saa:1999je}. 
Our main aim is to make a distinction between 
various types of chaotic behaviours. 
For our purpose, the comparison of the cases with different values
of $z_0$ may not be appropriate. Hence we analyze the cases with different 
values of $\alpha$, which may provide us continuous change
from a regular orbit to a very strongly chaotic one.

The parameters of particle orbits such as the energy and angular momentum
 are appropriately chosen such that the motion is bounded.
We choose the orbital parameters as
$H = - 0.2$ and $L = 1$,  and the disk width as 
$z_0 = 0.5$.
Figure~\ref{pmap-Newton} shows a set
 of Poincar\'{e} maps for different values of
the surface density ((a) $\alpha = 0.01$, (b) $\alpha = 0.1$, 
and (c) $\alpha = 10.0$). 
The equatorial 
plane $(z=0)$ is chosen for a Poincar\'{e} section. 
We plot the points on the 
$(\varpi,v^\varpi)$ plane  when the particle crosses the 
Poincar\'{e} section with $v^z >0$. In these figures, trajectories starting 
from various initial conditions are shown. From Figs.~\ref{pmap-Newton}
(a) and (c), 
we confirm these system are almost integrable. 
The outermost trajectories are 
called Orbit (a) and Orbit (c).
On the other hand, 
a widespread chaotic sea is found in Fig.~\ref{pmap-Newton}
(b). This is because the 
forces by the point mass and by the disk are comparable and those are
competing each other. 
In Fig. \ref{pmap-Newton} (b), we see 
the ``outermost" trajectory with the initial condition of
$(\varpi,v^\varpi,z,v^z)=(1.2,0,0,0.76)$ (called Orbit (b)) 
is not a simple torus but 
forms an almost two dimensional distribution
in which the orbital points are widely 
scattered. It means that
the particle motion is chaotic.

Figure~\ref{lyapu-Newton} shows the time evolution of the Lyapunov exponents 
for Orbits (a), (b), and (c) in Fig.~\ref{pmap-Newton}. 
Here we show a ``local" Lyapunov exponent defined in Appendix~\ref{loclyapu}. 
We only refer to the integration time interval  $t_\Delta$ to define it
(see Appendix in more details). $t_\Delta$ is chosen to be 
$t_\Delta =10^4$, which satisfies the condition of 
$t_D \ll t_\Delta \ll t_{T}$
with $t_D (\approx 10 \sim 10^2)$ and $t_{T}(\approx 10^6)$ being 
the dynamical time and the total integration period of our calculation,
 respectively. We also calculate it with other time intervals, 
$t_\Delta=2 $ or $4\times 10^4$.
We find that the result is not sensitive to this choice. 
We numerically calculate the exponents with the algorithm 
shown in \cite{Shimada} and show the maximum component of it.

The value for Orbit (a) is very small
and almost constant~\cite{foot}. 
For Orbit (c), 
the system is not exactly integrable. 
The system in the limit of
$\alpha\rightarrow \infty$
is of course integrable, 
but there is no bound orbit in such a limit.
Since we are analyzing a bound orbit, 
even if $\alpha$ is very large,
we cannot ignore the gravitational effect of a point mass.
It makes the system nonintegrable.
Nevertheless, the motion looks very regular
(see the Poincar\'{e} map in Fig.~\ref{pmap-Newton}).
In fact we find a very small Lyapunov exponent,
which is smaller 
than that of Orbit (a) as shown in Fig.~\ref{lyapu-Newton}.
This value is also almost constant, which means that
the strength of the chaos does not change much in time.
Hence we may regard this orbit as a regular one.

\begin{figure}
\includegraphics[width=7.5cm]{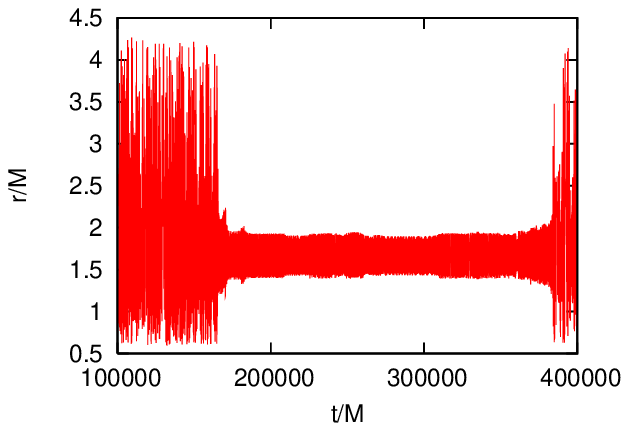}
\includegraphics[width=7.5cm]{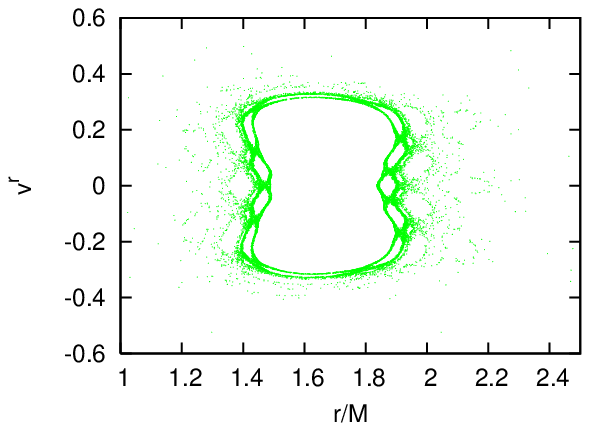}
\caption{\label{local-ana} The particle motion in the $r$-direction 
in terms of time and 
its Poincar\'{e} map for the time interval between 
$t/M = 10^5 $ and $t=4\times 10^5$ for Orbit (b).
The Poincar\'{e} map shows there exist many small tori around the origin.}
\end{figure}

\begin{figure}
\includegraphics[width=7.5cm]{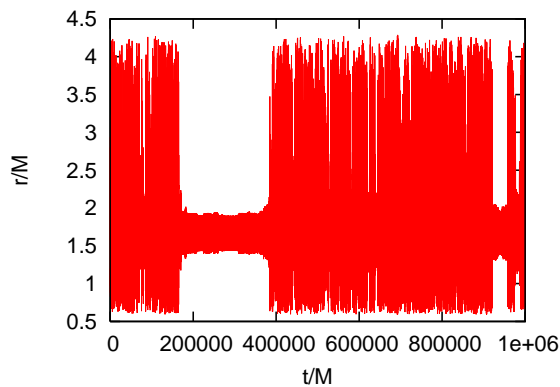}
\caption{\label{traj_long} The particle motion 
of Orbit (b) in the $r$-direction 
 for a longer time interval than that in Fig.~\ref{local-ana}.
There exists 
new stagnant motion as the same one
 in Fig.~\ref{local-ana}. }
\end{figure}

On the other hand 
Orbit (b) gives large positive Lyapunov exponent.
It also shows time variation. 
We should note that the value quickly goes down 
to a very small one in the time interval of
$t/M = (1.6 \sim 3.8) \times 10^5$. We pick up the 
data around this interval and show the time evolution of the 
$r$-position of the 
particle and the Poincar\'{e} map in Fig.~\ref{local-ana}.
From this, 
we find that although 
the particle motion in Orbit (b) is chaotic, it
stays around $r \sim 1.2-2.2 M$ in the time interval
of $t/M=(1.6\sim 3.8) \times 10^5$.
The motion in this period seems to be nearly regular.
In fact, the ``local" 
Lyapunov exponent decreases to $5 \sim 6 \times 10^{-3}$, 
which is almost 
the same as those of Orbits (a) and (c).
We call this phase of motion Orbit (b-2).
The phase before this interval, in which a particle motion looks
very chaotic, is called Orbit (b-1).
We have also performed  numerical integration
for longer time period and confirm 
such phases as Orbit (b-2)
often appears in a chaotic orbit
(see  Fig.~\ref{traj_long}).
The important point is that
two different phases of motion appear and 
both a nearly integrable and a 
more strongly chaotic motion co exist in the 
same trajectory. 

\begin{figure}
\includegraphics[width=7.5cm]{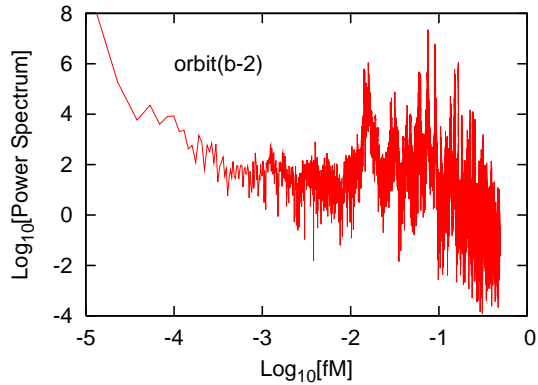}
\caption{\label{power_B} The power spectrum of the motion in 
the $r$-direction of
Orbit (b-2), which is part of nearly regular motion in
Orbit (b) ( $t/M=(1.6\sim 3.8) \times 10^5$). 
We find a power-law spectrum, e.g., a $1/f$ fluctuation.
This is a reflects the existence of small tori 
in the phase space, and the particle moves almost regularly  
there~\cite{Koyama:2007cj}.}
\end{figure}

The Poincar\'{e} map of Orbit  (b-2) in Fig. \ref{local-ana}
shows that many small tori exist.
 It is well known that 
such a structure appears if an orbit is nearly integrable and 
produces the so-called $1/f$ fluctuation~
\cite{Karney:1995,Contopoulos:2000,Contopoulos:2002,Koyama:2007cj}.
Then we also analyze the power spectrum of the $r$-component of 
Orbit (b-2), which 
clearly shows a $1/f$ fluctuation 
for $f\leq 10^{-2} M^{-1}$
(see Fig.~\ref{power_B}).
This confirms our previous result~\cite{Koyama:2007cj}
 in the present model.

\subsection{Indication of chaos in gravitational waves}
Next we study how to extract information from  
such a chaotic system and distinguish the orbits,
i.e., a nearly integrable and more strongly chaotic motions.

\begin{figure}
\includegraphics[width=7.5cm]{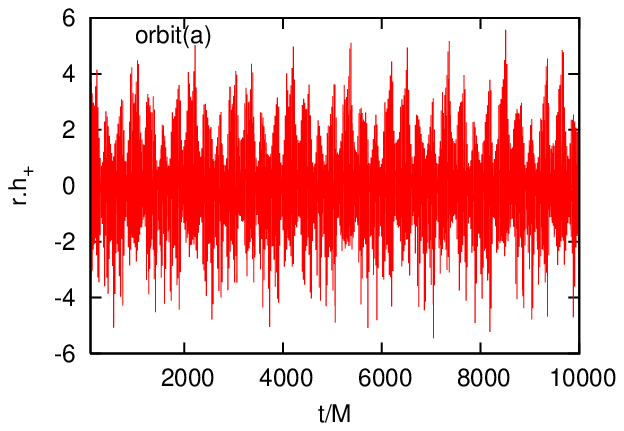}
\includegraphics[width=7.5cm]{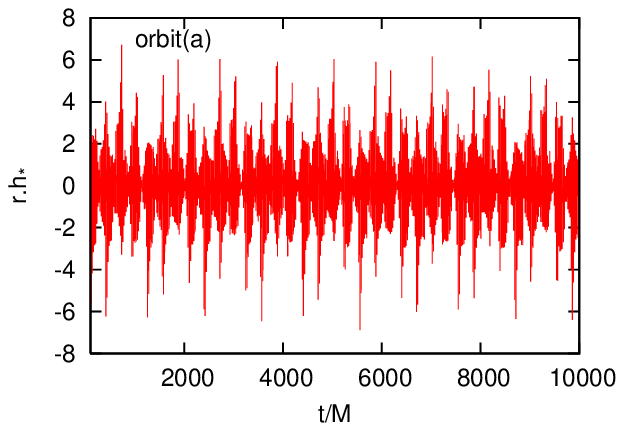}
\includegraphics[width=7.5cm]{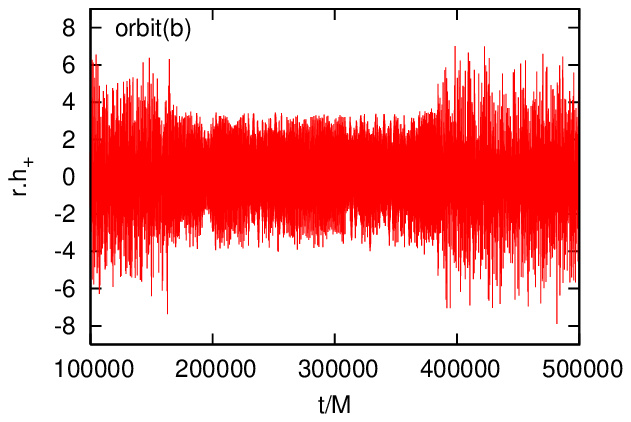}
\includegraphics[width=7.5cm]{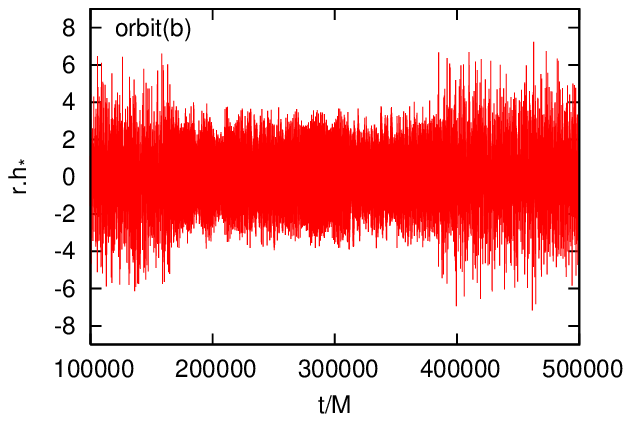}
\includegraphics[width=7.5cm]{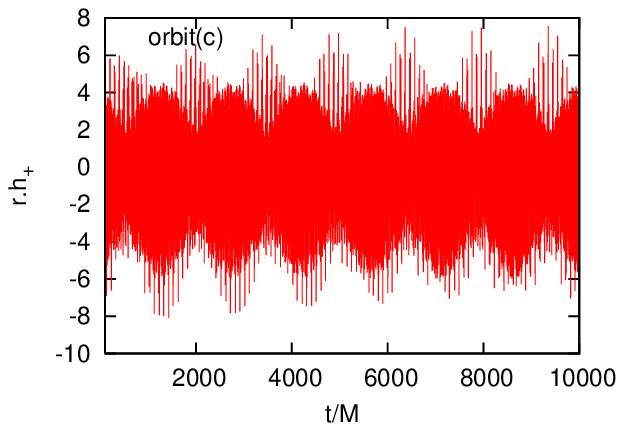}
\includegraphics[width=7.5cm]{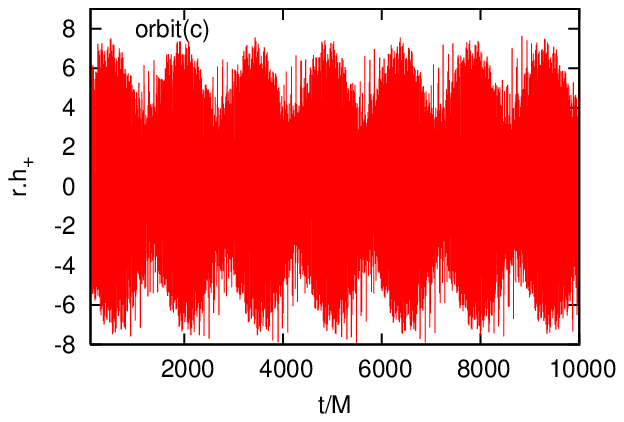}
\caption{\label{waveform-Newton} The gravitational waveforms 
evaluated by the quadrupole 
formula. 
Top, middle, and bottom figures correspond to those for Orbits~(a), (b), and 
(c),
respectively. The left and right rows give the  ``$+$'' and ``$\times$'' 
polarization modes, 
respectively.}
\end{figure}

In \cite{Koyama:2007cj}, the 
authors focused on the power spectrum of particle 
motion moving in Schwarzschild spacetime and found that
it shows a power-law behaviour.
 In this work, we use a similar analysis for the
gravitational waves emitted from our system.
It could be a new and robust way to
observe chaotic behaviors in astrophysical objects, 
as mentioned in our Introduction. 
The gravitational waves from the present system are 
calculated by the quadrupole formula~\cite{Landau},
which is given by
\begin{eqnarray}
&&h_+ = 
\left[\left(h^Q_{xx}-h^Q_{yy}\right)\cos 2\varphi
+2h^Q_{xy}\sin 2\varphi\right]\frac{(\cos^2\theta+1)}{4}
- \left(h^Q_{xx}+h^Q_{yy}-2h^Q_{zz}\right)
\frac{\sin^2\theta}{4}
 \nonumber\\
&&~~~~~~-\left(h^Q_{xz}\cos\varphi+h^Q_{yz}\sin\varphi
\right)\sin\theta\cos\theta
\,, \label{quad}\\
&&h_\times = 
\left[2h^Q_{xy}\cos2\varphi-(h^Q_{xx}-h^Q_{yy})\sin2\varphi\right]
\frac{\cos\theta}{2}
+\left(h^Q_{xz}\sin\varphi-h^Q_{yz}
\cos\varphi\right)\sin\theta, 
\label{quad2}
\end{eqnarray}
where
\begin{eqnarray}
h_{ij}^Q \equiv  \frac{2}{r}\frac{d^2 Q_{ij}}{dt^2}
~~~~{\rm with}~~
Q_{ij} \equiv  \mu\left(  Z^i Z^j - \frac{1}{3}\delta_{ij}~\vect{Z}^2 
\right)
~~({\rm the~reduced~quadrupole~moment~of ~a~point~mass}).
\end{eqnarray}
$(r,\theta,\varphi)$ [or $(x,y,z)$]
 is the position of a distant observer in spherical coordinates [or 
Cartesian coordinates], and $\vect{Z}(t)$ is a trajectory
 of a particle.
We assume that the observer is on the equatorial plane, i.e.
$(\theta,\varphi)=(\pi/2,0)$.
 Figure~\ref{waveform-Newton} shows the waveforms from Orbits (a), (b), 
and (c). The left panels show the ``$+$'' polarization modes
of those waves, while  
the right ones are the 
``$\times$'' polarization. 
The top, middle, and bottom panels correspond 
to the waves from Orbits (a), (b), and (c), respectively.
The waves from Orbits (a) and (c) show a periodic feature, 
which is expected 
from the Poincar\'{e} maps in Fig.~\ref{pmap-Newton}. On the other hand, the 
waves from Orbit (b) show a completely different behaviour. 
We find much random spiky noise in the waveform
before $t/M =  1.6 \times 10^5$
and after $t/M = 3.8\times 10^5$.
This is a typical feature of the gravitational waves
from highly chaotic motion~\cite{Kiuchi:2004bv,Suzuki}.
We also find that the amplitude  
decreases for the time interval of
 $t/M =  (1.6 \sim 3.8) \times 10^5$. As shown in 
Fig.~\ref{local-ana}, in this time interval,
 the particle moves near the small tori
in the phase space. This adjective feature of this 
 particle motion appears clearly 
in the gravitational amplitudes.
That is, in the phase of a nearly regular motion,
 the particle position and its velocity do not change much 
compared with those in the more strongly chaotic phase (b-1) 
(see Fig.~\ref{pmap-Newton}(b) and Fig.~\ref{local-ana}(b)). 
The time variation of the 
quadrupole moment of the system is small and hence the 
wave amplitude decreases as well.

\begin{figure}
\includegraphics[width=7.5cm]{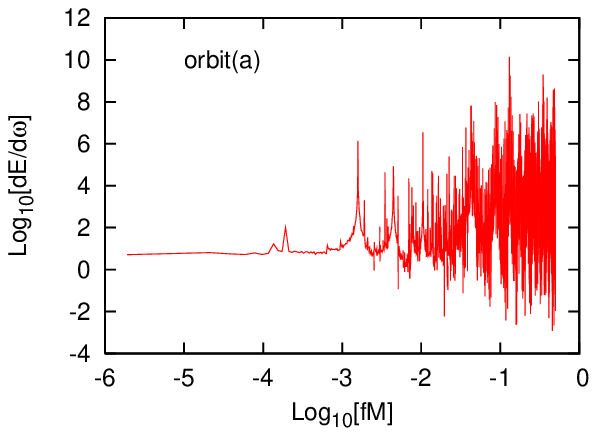}
\includegraphics[width=7.5cm]{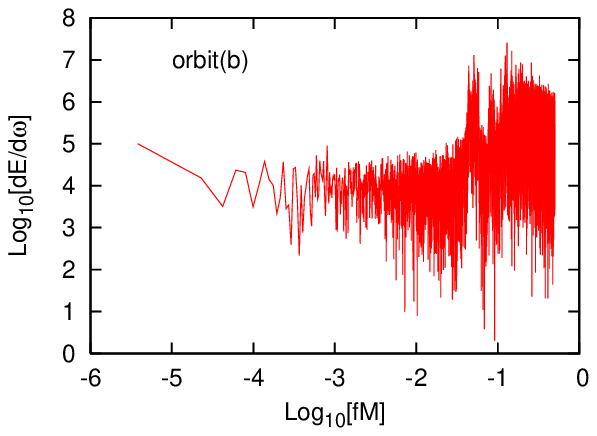}
\includegraphics[width=7.5cm]{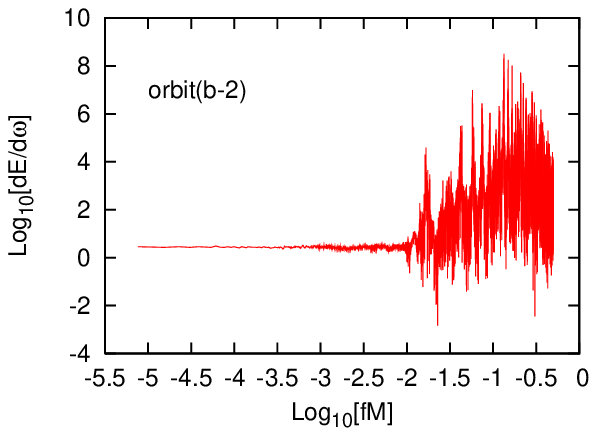}
\includegraphics[width=7.5cm]{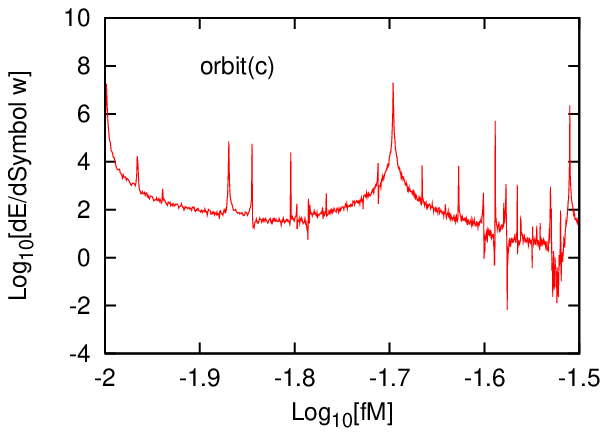}
\caption{\label{energyspec-Newton} The 
energy spectra of the gravitational waves shown
in Fig.~\ref{waveform-Newton}.
Orbit (b-2) gives the spectrum of the waves for the ``stagnant motion'',
i.e., when the particle motion of Orbit (b) becomes near regular
for  
$t/M = (1.6  \sim 3.8)\times 10^5$. Figures~(a) and (c) show many
sharp peaks at certain 
characteristic frequencies. This is because of the regular motion. 
The spectrum 
in Fig.~(b), which looks like white noise
for $fM \leq 10^{-2}$, is clearly different from those 
in Figs.~(a) and (c),
but the spectrum in Fig. (b-2) does not look like white noise. 
It looks similar to the cases (a) and (c).
However, the peaks are not sharp but rather 
broadened by appearing so many other spikes.
Note that the typical frequency of the orbits is in the range of
$fM = 10^{-2} \sim 10^{-1} $ (see Fig.~\ref{power_B}). 
 }
\end{figure}

We also calculate the energy spectra of the gravitational waves,
which will be one of the most important observable quantities
in the near future.
In Fig.~\ref{energyspec-Newton}, we show the 
energy spectra for  each orbit.
Figures~\ref{energyspec-Newton}(a) and (c) 
show many sharp peaks at certain characteristic frequencies. 
If a motion is regular, we expect several typical frequencies
with those harmonics.
So such a result reflects that the particle moves regularly.
Figure~\ref{energyspec-Newton} (b) gives
 the spectrum of Orbit (b).  It
is clearly different from the previous two almost 
regular cases.
It looks just like white noise,
below a typical frequency $fM \sim 10^{-2} $, i.e.,
the shape of the spectrum is flat and it 
contains many noisy components.
However, 
the spectrum of Orbit (b-2) (Fig.~\ref{energyspec-Newton}(b-2)),
which is analyzed by the orbit only 
in the time interval of $t/M = (1.6  \sim 3.8)\times 10^5$, 
does not do so.  Rather it looks similar to 
 the spectrum of a regular orbit. Contrary to Fig.~\ref{energyspec-Newton}(b), 
it does not contain much noise at the low frequency region
($fM\leq 10^{-2}$).

\begin{figure}
\includegraphics[width=7.5cm]{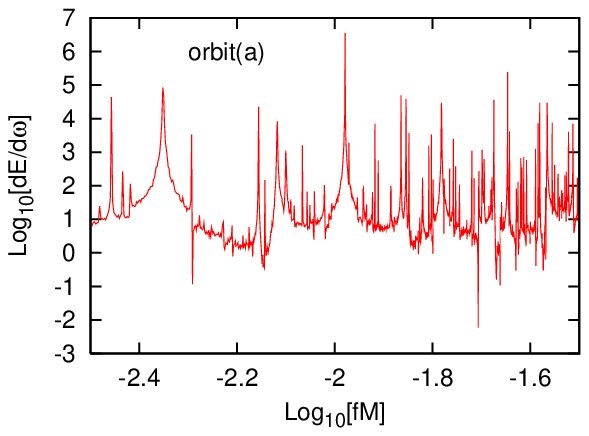}
\includegraphics[width=7.5cm]{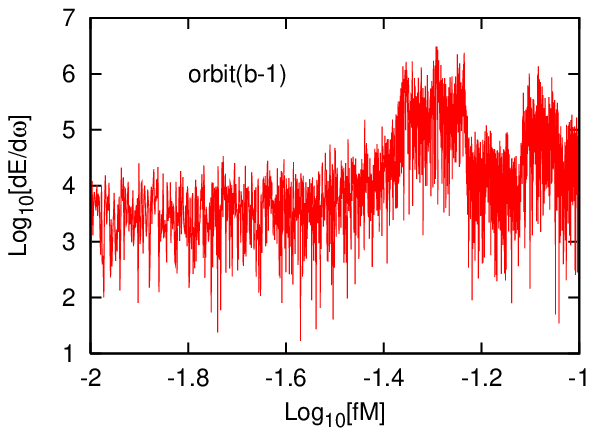}
\includegraphics[width=7.5cm]{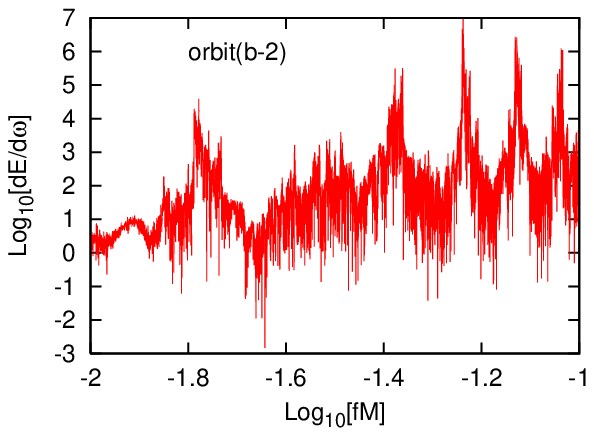}
\includegraphics[width=7.5cm]{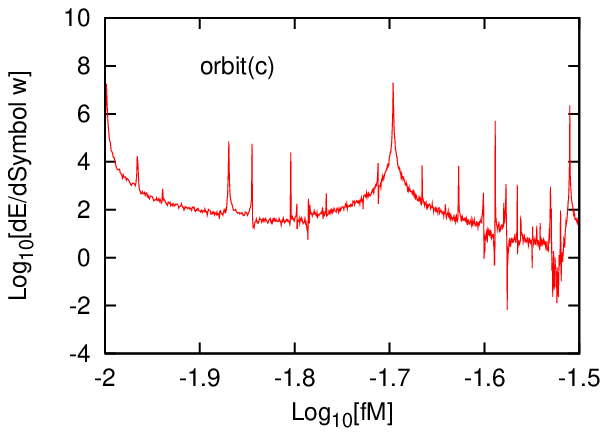}
\caption{\label{energyspec-Newton-Mag} Magnification of energy spectra of 
Orbits (a), (b-1), (b-2), and (c).}
\end{figure}

To see more detail,  
dividing the time interval of Orbit (b) into two,
we show  the magnifications of the 
spectra of Orbits (a), (b-1), (b-2), and (c)
in Fig.~\ref{energyspec-Newton-Mag}. 
Compared to the spectra (a) and (c), 
the spectra (b-1) and (b-2)  contain 
many noisy spikes.
Such noisy spikes are usually 
found in the gravitational waves from
 a chaotic orbit~\cite{Kiuchi:2004bv}. 
However, the spectra (b-1) and (b-2) are completely
different. The spectrum (b-1)
is just white noise.
No structure is found.
On the other hand, the spectrum (b-2) looks similar to 
those for regular orbits.
The ``sharp" peaks appear at some frequencies, but
 the widths of those peaks are broadened
by many noisy spikes. 
Therefore, we conclude that Orbit (b-2)
looks nearly regular but still holds its chaotic character, 
and such a feature imprints in the spectrum of the waves.
The important point is that two phases
 in the particle orbit (b), i.e., the nearly regular phase
and the more strongly chaotic one, 
are  also distinguishable 
in the gravitational wave forms and the energy spectra. 
With this analysis, we could constrain 
orbital parameters.


\section{Summary and Discussion}\label{summary}
In this paper we have investigated chaos characteristic for a test 
particle motion in a system of a point mass with a massive 
disk in Newtonian gravity. 
To distinguish such characteristics, we propose the 
gravitational waves emitted from this system.
 At first, we analyzed the 
motion of the particle by use of 
the Poincar\'{e} map and the ``local" Lyapunov exponent.
We found that the phase in which particle motion becomes 
 nearly regular always 
appears even though the global motion is chaotic.
We emphasize that 
both phases of  nearly regular and more strongly chaotic motions
are found in the same orbit. 

The gravitational wave forms and their
 energy spectra have been evaluated by use of the quadrupole formula
 in each case. In two almost regular cases, the waves show the periodic 
behaviour and certain sharp peaks appear in those energy spectra. 
In the chaotic case, we have found that
 the waves show two phases, 
the nearly regular phase and a more strongly 
chaotic one.
In the nearly regular phase, wave amplitude
gets smaller in the more strongly 
chaotic phase. 
The energy spectra are also clearly different. 
The spectrum in the more strongly chaotic
phase looks like white noise, but in the nearly regular one, it becomes
similar to those in the regular ones.
However 
it is accompanied by many small noisy spikes, which is a characteristic
feature of a chaotic system.
These spikes make the widths of the 
spectrum peaks broader 
than those in the regular cases.
Comparing information from the waves with 
the particle motion, we conclude that we can extract 
chaotic characteristics of a particle motion of 
the gravitational waves of the system.
In the present analysis, in the spectrum (b-2) of the gravitational waves,
we do not find a power-law structure,
which appears in the spectrum of the particle motion. 
This may be because the waveform is given by the change of the quadrupole
moment, which contains  higher time derivatives of a particle trajectory
such as acceleration.
It may be much more interesting if one can
find the $1/f$ behaviour in some information of
the gravitational waves
because such an indication may specify the type of chaos more clearly.
This is under investigation.

Finally, we mention a possibility to constrain parameters
in a dynamical system. 
If the gravitational waves are observed for a sufficiently
 long time, we can monitor the time variation of the wave amplitudes,
their forms and polarizations.
We can then calculate the energy spectra 
for some durations. 
If the spectra show one of the typical 
characteristics found in this paper, the 
parameters of a particle motion could be constrained.
Of course, a realistic system can be more complicated, and the 
present model may be too simple.
But we believe the characteristic behaviour of the gravitational waves 
found in this paper 
will help us to understand a chaotic system. 
Therefore our next task is to 
analyze the gravitational waves from various chaotic systems, especially 
relativistic chaotic systems~\cite{Lemos:1993qp,Karas,Varvoglis,Bombelli,
Moeckel,Yust,Sota,Suzuki,letelier_viera,podolsky_vesely,
Kiuchi:2004bv,Koyama:2007cj}.
 Then, we should investigate whether  or not
the correlation between the gravitational waves and chaos 
in dynamical systems found 
in this work is generic.

\section*{Acknowledgments}

We express thanks to T. Konishi for useful discussions.
This work was supported in part by Japan Society for 
Promotion of Science (JSPS) Research Fellowships (K.K. and H.K.), 
by a Grant-in-Aid from the Scientific
Research Fund of the JSPS (No. 17540268), and by the
Japan-U.K. Research Cooperative Program.
K.M. would like to thank DAMTP, the Centre for Theoretical Cosmology,
and Clare Hall, where this work was completed.



\appendix
\section{Local Lyapunov Exponent}\label{loclyapu}
In this appendix, we give the 
definition of ``local'' Lyapunov exponent. 
Our definition of 
``local'' Lyapunov exponent is somewhat different 
from the conventional one\cite{LLE},
but those are essentially the same.

At first, let us consider the system whose time evolution is described 
by a set of differential equations in $N$-dimensional space,
\begin{eqnarray}
\dot{{\bf x}} = {\bf F}({\bf x})
\,,
\end{eqnarray}
where ${\bf x}(t)$ is a $N$-dimensional vector.

The time evolution of the orbital deviation $\delta {\bf x}$,
which is the difference between two nearby orbits, obeys 
the following set of linear differential equations:
\begin{eqnarray}
\delta \dot{{\bf x}} = \frac{\partial {\bf F}}{\partial {\bf x}}({{\bf x}(t)})
\delta {\bf x}. \label{linear}
\end{eqnarray}
The solution of Eq.~(\ref{linear}) can be written formally as 
\begin{eqnarray}
\delta {\bf x}(t) = U^t_{~t_0} \delta {\bf x}_0,
\end{eqnarray}
where $\delta {\bf x}_0$ is an ``initial" deviation at some time $t_0$ and 
$U^t_{~t_0}$ is an evolution matrix, which is given by the following
integration;
\begin{eqnarray}
U^t_{~t_0} = \exp \left[ \int^t_{t_0} 
\frac{\partial {\bf F}}{\partial {\bf x}}
({{\bf x}(t')})dt' \right]
\,.
\end{eqnarray}

We define the ``local'' Lyapunov exponent in time interval $[t_0,t]$ by 
\begin{eqnarray}
\lambda(e^k,t) = \frac{1}{t-t_0}\log 
\frac{||U^t_{~t_0} {\bf e}_1 \wedge 
U^t_{~t_0} {\bf e}_2 \wedge \cdots
\wedge U^t_{~t_0} {\bf e}_k
||}{||{\bf e}_1 \wedge {\bf e}_2 \cdots \wedge {\bf e}_k ||} \label{locdef}
\end{eqnarray}
for $k=1,2,\cdots , N$, where  $e^k$ is a $k$-dimensional subspace 
in the tangent 
space at the initial point ${\bf x}_0$, which is spanned by $k$ 
independent vectors ${{\bf e}_i}$ $(i=1,2,\cdots,k)$, 
$\wedge$ is an exterior product, and $||\circ||$ is a norm 
with respect to some appropriate Riemannian metric. 
If we take a limit of $t\rightarrow \infty$, $\lambda(e^k,\infty)$
correspond to the conventional Lyapunov exponents.

If the integration 
time interval $t_{\Delta}\equiv t-t_0$ is much longer than the 
dynamical time of 
the system, we may find convergent values for each $\lambda(e^k,t)$,
which are almost independent of $t_{\Delta}$ (or $t_0$).
We may call them ``local" Lyapunov exponents at $t$.
The maximum value of ``local" Lyapunov exponents, i.e.
$\lambda(t)=\max \{\lambda(e^k,t)|k=1,2,\cdots , N\}$
is the most important one for our discussion.
So we also call it the ``local" Lyapunov exponent at $t$.


\begin{thebibliography}{99}


\bibitem{Hobill}
{\em Deterministic Chaos in General Relativity}, edited by D. Hobill, A. Burd,
and A. Coley (Plenum, New York, 1994), and references therein.

\bibitem{Barrow}
J.D.~Barrow, Phys. Rep. {\bf 85}, 1 (1982).

\bibitem{Conto}
G.~Contopoulos, Proc. R. Soc. London {\bf A431},183(1990).

\bibitem{Karas}
V.~Karas and D.~Vokrouhlick\'{y}, Gen. Rela. Grav. {\bf 24},729(1992).

\bibitem{Varvoglis}
H.~Varvoglis and D.~Papadopoulos, Astron. Astrophys. {\bf 261},664(1992).

\bibitem{Bombelli}
L.~Bombelli and E.~Calzetta,  Class. Quantum Grav. {\bf 9},2573(1992).

\bibitem{Moeckel}
R.~Moeckel, Commun. Math. Phys. {\bf 150},415(1992).

\bibitem{Dettmann}
C.P.~Dettmann, N.E.~Frankel and N.J.~Cornish, Phys.\ Rev.\ D {\bf 50},
618(1994).

\bibitem{Yust}
U.~Yurtsever, Phys.\ Rev.\ D {\bf 52},3176(1995).

\bibitem{Sota}
Y. Sota, S. Suzuki, and K. Maeda, Class. Quantum Grav. \textbf{13},1241(1996).

\bibitem{Suzuki}
S. Suzuki and K. Maeda, Phys. Rev. D \textbf{55},4848(1997); 
 \textbf{58},023005(1998);
 \textbf{61},024005(1999).

\bibitem{letelier_viera} P.S. Letelier and W.M. Viera, Phys. Rev.
D {\bf 56},8095(1997).

\bibitem{podolsky_vesely}  J. Podolsky and K. Vesely, Phys. Rev. D {\bf 58},
081501(1998).


\bibitem{Levin}
J.~Levin, Phys. Rev. Lett. {\bf 84},3515(2000).

\bibitem{Schnittman}
J.D.~Schnittman and F.A.~Rasio, Phys.\ Rev.\ Lett.\  {\bf 87}, 121101 (2001).

\bibitem{Cornish}
N.J.~Cornish and J.~Levin, Phys. Rev. Lett. {\bf 89},179001(2002).


\bibitem{Kiuchi:2004bv}
  K.~Kiuchi and K.~Maeda,
  Phys.\ Rev.\  D {\bf 70},064036(2004)


\bibitem{Karney:1995}
 C.F.F. Karney,
 Physica~D {\bf 8},360(1983)

\bibitem{Contopoulos:2000}
 G.~Contopoulos, M.~Harsoula and N.~Voglis,
 Celest.\ Mech.\ Dyn.\ Astron.\ {\bf 78},197(2000)

\bibitem{Contopoulos:2002}
G.~Contopoulos, {\em Order and Chaos in Dynamical Astronomy},  (Springer, 2002)

\bibitem{Koyama:2007cj}
  H.~Koyama, K.~Kiuchi and T.~Konishi,
  arXiv:gr-qc/0702072.
 

\bibitem{Tsubono:1994sg}
  K.~Tsubono,
{\it Prepared for Edoardo Amaldi Meeting on Gravitational Wave Experiments,
 Rome, Italy, 14-17 Jun 1994}

\bibitem{Abramovici:1992ah}
  A.~Abramovici {\it et al.},
  ``LIGO: The Laser interferometer gravitational wave observatory,''
  Science {\bf 256}, 325 (1992).

\bibitem{Hough:1996nx}
  J.~Hough {\it et al.},
{\it Prepared for TAMA Workshop on Gravitational Wave Detection, Saitama,
 Japan, 12-14 Nov 1996}

\bibitem{Thorne:1995xs}
  K.S.~Thorne,
   arXiv:gr-qc/9506086.

\bibitem{Saa:1999je}
  A.~Saa and R.~Venegeroles,
  Phys.\ Lett.\ A {\bf 259},201(1999)

\bibitem{Saa:2000ec}
  A.~Saa,
  Phys.\ Lett.\ A {\bf 269},204(2000)

\bibitem{footnote1}
The recent observation suggests there exist huge black holes 
at the centers of many galaxies.
See, for example,~\\
  J.~Kormendy and D.~Richstone, Astrophys.\ J.\  {\bf 393},559(1992)


\bibitem{Shimada}
 I.~Shimada and T.~Nagashima, PTP {\bf 61}, 1605 (1979)

\bibitem{foot}
The reason why we find non-zero positive Lyapunov exponent 
for an integrable system is that 
we solve the equation of motion by a finite difference method and
the finite difference approximation does not provide an 
exact integrable system.
In fact, if we reduce the time step for integration,
the value decreases.


\bibitem{Saslaw:1985}
  W.C.Saslaw,~{\it Gravitational physics of stellar and galactic systems},
  Cambridge University Press, 1985.

\bibitem{Lemos:1993qp}
  J.P.S.~Lemos and P.S.~Letelier,
  Phys.\ Rev.\ D {\bf 49},5135(1994).

\bibitem{Landau} L. D. Landau and E. M. Lifshitz, {\it The Classical Theory
of Fields}, (Pergamon, Oxford, 1951).


\bibitem{LLE}
P.~Grassberger, R.~Badii, and A.~Politi, J.~Stat.~Phys.~{\bf 51},~135 (1988);
H.~E.~Kandrup,~B.~L.~Eckstein, and B.~O.~Bradley, Astron.~Asrophys.~{\bf 320},
 65 (1997)

\end{thebibliography}
\end{document}